\documentstyle[twocolumn,aps,epsf,floats]{revtex}
\begin{document}

\twocolumn[\hsize\textwidth\columnwidth\hsize\csname %
@twocolumnfalse\endcsname

\title{Probing the susceptibility $\chi'({\bf q})$ in cuprates using Ni
impurities}
\author{Dirk K. Morr, J\"{o}rg Schmalian, Raivo Stern, Charles P. Slichter}
\address{Department of Physics and Science and Technology Center for
Superconductivity, University of Illinois at Urbana-Champaign, 1110
W. Green Street, Urbana, IL 61801}
\date{\today}
\maketitle
\begin{abstract}
Recent NMR experiments on YBa$_2$(Cu$_{1-x}$Ni$_x$)$_3$O$_{6+\delta}$
(Bobroff {\it et al.} Phys. Rev. Lett. {\bf 79}, 2117 (1997)) have shown
that Ni impurities provide an important probe for the momentum
dependence of the static spin susceptibility $\chi'({\bf q})$.
Combining the
measurements of  the $^{17}$O line width $\Delta \nu$ with those of the Gaussian relaxation time
$T_{\rm 2G}$,
we find that the magnetic correlation length $\xi$ must have a substantial
temperature dependence.
 Using numerical
simulations we present a detailed analysis of  $\Delta \nu$
as a function of temperature and impurity concentration. 
 For a Lorentzian shape of
$\chi'({\bf q})$ $\Delta \nu$ strongly depends on $\xi$, in contrast to the
Gaussian form. This result together with the experimental finding that $\xi$
is temperature dependent rules out the  Gaussian form of $\chi'({\bf q})$.
\end{abstract}
 ]

\narrowtext
Precise knowledge  of the spin susceptibility $\chi({\bf q}, \omega)$ of the
cuprates is essential for understanding their unusual normal state properties.
The imaginary part,  $\chi^{\prime \prime}({\bf q},\omega)$ can be probed
either
by inelastic neutron scattering (INS) \cite{Kei,Bou,Hay,Mook97}, or in the low
frequency limit by NMR measurements of the spin-lattice relaxation rate
$1/T_1$ \cite{T1}. In  contrast, one knows little about the real part
of the susceptibility, $\chi^\prime({\bf q})$,  since information can, so
far,
only be extracted from the NMR observation of the Gaussian component of the
transverse relaxation time, $T_{\rm 2G}$, of planar Cu~\cite{PS91,Curro97}.
In particular, the analysis of INS and NMR experiments has not yet led to a
consensus on the shape of $\chi({\bf q},\omega)$  in momentum space and the
temperature ($T$) dependence of the antiferromagnetic correlation length,
$\xi$.
In this communication we present new insight into this issue based on
experiments by Bobroff {\em et al.}~\cite{BAY97}. Our principal conclusions
are
that $\xi$ 
in YBa$_2$Cu$_3$O$_{6+\delta}$ is $T$-dependent and that the Lorentzian
form  of $\chi^\prime(\bf q)$ provides a completely consistent description of
the data,  whereas the  Gaussian form can be ruled out.

Bobroff {\em et al.}~\cite{BAY97} recently presented a novel approach to the
measurement of $\chi^\prime({\bf q})$ using Ni impurities in
YBa$_2$(Cu$_{1-x}$Ni$_x$)$_3$O$_{6+\delta}$. These impurities induce a spin
polarization at the planar Cu sites via $\chi^\prime({\bf q})$.  The
hyperfine coupling between Cu and O induces  a spatially varying
polarization and  an additional  broadening
\begin{equation}
\Delta \nu_{\rm imp} = \Delta \nu-\Delta \nu_0=\alpha f(\xi)/T
\label{dnu}
\end{equation}
of the planar $^{17}$O NMR, where $\Delta \nu$ and $\Delta \nu_0$
are the total and $x=0$ line width, respectively.
In Eq.~\ref{dnu} $\alpha$ is the overall amplitude of $\chi^\prime({\bf q})$ and $f(\xi)$
characterizes the dependency of $\Delta \nu$ on $\xi$ ($\alpha=4\pi \chi^*$ in
the notation of Ref.~\cite{BAY97} and\cite{Morr97}). Finally, the factor
$1/T$
is caused by the Curie behavior of the  Ni impurities in
YBa$_2$Cu$_3$O$_{6+\delta}$\cite{Mah94,Men94} with effective moment
$p_{\rm eff}\approx 1.9 \mu_{\rm B}$ ($1.59 \mu_{\rm B}$) for $\delta=0.6
(\delta=1)$.
Bobroff {\it et al.} found that $T \Delta \nu(T)$  strongly depends on
temperature
and  the Ni concentration $x$ in the sample.  Furthermore they observed a much
stronger broadening in the underdoped, $\delta=0.6$, sample than in the
overdoped
one with $\delta=1$. Performing numerical simulations of the NMR line shape by
assuming a Gaussian form for  $\chi^\prime({\bf q})$, they found that $f(\xi)$
is basically constant for all physically reasonable values of $\xi$.
Combining these results with $T_{\rm2G}$ data by Takigawa \cite{Tak94}, they
concluded that  $\xi$ is $T$-independent for the underdoped samples.
On the other hand, in every scenario of cuprate superconductors in which the
anomalous low-energy behavior  is driven by spin fluctuations one would expect
the correlation length $\xi$ to be $T$-dependent
 (for recent reviews, see: \cite{Pines,Scal}).
Thus their result has important implications about the mechanism of
superconductivity. We recently pointed out \cite{Morr97}, that our
simulations using a Lorentzian form of $\chi^\prime({\bf q})$ yield a
different
result and are actually compatible with a $T$-dependent $\xi$.

\begin{figure} [t]
\begin{center}
\leavevmode
\epsfxsize=7.5cm
\epsffile{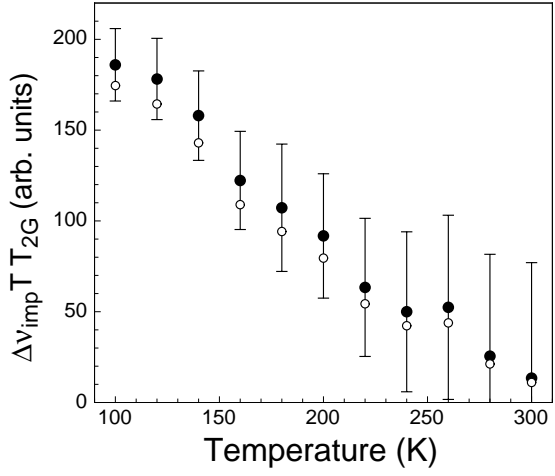}
\end{center}
\caption{ $T \Delta \nu_{\rm imp} T_{\rm 2G} $ as a function of temperature, using the original (empty circles) or $T_1$ corrected (filled circles) $T_{\rm 2G}$ data \protect\cite{T1corr}.                
This product is independent of $\alpha$ and its $T$-dependence
must be caused by a $T$-dependent $\xi$.
We used the data from Ref.~\protect\cite{BAY97} and \protect\cite{Tak94}. }
\label{prod}
\end{figure}

Before going into the details of our calculations, it is important to
notice that the
fact that $\xi$ must be $T$-dependent can be deduced even without a
detailed model
from the very experimental data by Bobroff {\em et al.}~\cite{BAY97} for
$\Delta \nu(T)$ and Takigawa \cite{Tak94} for $T_{\rm 2G}$. To show this,
we need
to recognize that we can always express $T_{\rm 2G}$ as a product of
$\alpha$ and
a function of $\xi$, namely
\begin{equation}
T_{\rm 2G}^{-1} = \alpha g(\xi) \ .
\label{t2g}
\end{equation}
We can then eliminate $\alpha$ by forming the product
\begin{equation}
T \Delta \nu_{\rm imp} T_{\rm 2G} = { f(\xi) \over g(\xi)}
\label{product}
\end{equation}
which depends solely on $\xi$.
In Fig.~\ref{prod}, we plot the  product $T_{\rm 2G} T \Delta \nu_{\rm imp}$
as a function of $T$\cite{T1corr}.
We see that this product is strongly $T$-dependent, dropping by more than a
factor
of 2 between $100 \, K$ and $200 \, K$. Therefore $\xi$ must
 have a substantial
 $T$-dependence.

To  have a more quantitative insight into the $T$-dependence of
$T \Delta \nu(T)$ of Ref.~\cite{BAY97},  we must go into details.
We  present in the following  a theoretical analysis of the $^{17}$O line shape
using a method first applied by Bobroff {\it et al.}, to simulate
their experimental data.

To simulate  the $^{17}$O line shape numerically, we  distribute Ni impurities
on a $(100 \times 100)$ lattice with concentration $\frac{3}{2}x$ 
randomly at positions ${\bf r}_j$
on a two dimensional lattice \cite{com2}. We consider the Ni impurities as
foreign
atoms embedded in the pure material, which is characterized by a non-local
spin-susceptibility $\chi'({\bf q})$. In the following ${\bf s}_i$ characterizes the spin
dynamics of
the pure material,  ${\bf S}_i$ the difference at the
Ni site brought about by the Ni. These Ni spins polarize the spin ${\bf
s}_j$ of the
itinerant strongly correlated electrons.  To calculate the induced
moments
we need to know how the Ni impurities couple to these spins. Without
discussing the
microscopic origin of the effective Ni spin ${\bf S}_j$,  we assume that it
obeys a
Curie law, and that the coupling to the  spin ${\bf s}_j$ occurs via an
on-site
interaction described by
\begin{equation}
{ \cal H}_{int} = -J \sum_{j}  {\bf s}_j \cdot {\bf S}_j \  .
\label{Hint}
\end{equation}
The coupling constant $J$ is an unknown parameter of the theory and will be
estimated
below. Furthermore, we will assume like Bobroff {\it et al.}~that the Ni
impurities
do {\it not} change the magnetic correlation length or the magnitude of the
spin
susceptibility.

For the NMR experiments we consider an external magnetic field $B_0$ along
the $z$-direction. The Ni spins have a non-zero average value
obeying
$\langle S^z_j \rangle = C_{\rm Curie} B_0/T $ with Curie constant
$C_{\rm Curie}=p_{\rm eff}/(2\sqrt{3} k_{\rm B})$~\cite{Mah94,Men94}.

Adopting a mean field picture, the induced polarization   for
the electron spins at the Cu sites ${\bf r}_i$ is given by
\begin{equation}
\langle s^z_i \rangle  = \frac{J}{(g\mu_{\rm B})^2} \sum_j
\chi'({\bf r}_i-{\bf r}_j)  \langle S^z_j \rangle\, .
\label{Sind}
\end{equation}
Here, $\chi'({\bf r})$ is the real space Fourier transform of $\chi'({\bf
q})$. In the
following we consider two different forms of the spin susceptibility~\cite{mp92}. For the
commensurate case, there is only one peak, whereas in the incommensurate
case, one
has to sum over four peaks. The Gaussian form of $\chi'({\bf q})$ is given by
\begin{equation}
\chi_{\rm G}'({\bf q})=\alpha \xi^2 \exp\left(-({\bf q}-{\bf Q})^2
\xi^2\right)
\label{chig}
\end{equation}
and the Lorentzian form by
\begin{equation}
\chi_{\rm L}'({\bf q})=\alpha \xi^2/(1+({\bf q}-{\bf Q})^2 \xi^2) \ .
\label{chil}
\end{equation}
Since the question whether there exist incommensurate peaks in
YBa$_2$Cu$_3$O$_{6+\delta}$ has not been settled yet \cite{Mook97}, we will
consider
below both cases, a commensurate wavevector ${\bf Q}=(\pm \pi, \pm \pi)$,
and an incommensurate one with ${\bf Q}=\delta_{\rm i} (\pm \pi, \pm \pi)$. The
calculation
of the real space Fourier transform  finally  yields
\begin{eqnarray}
\chi_{\rm G}'({\bf r}) &=&\frac{\alpha}{4\pi} F({\bf Q})  \exp
 \Big( - {  {\bf r}^2 \over 4 \xi^2} \Big) \, ,\nonumber \\
\chi_{\rm L}'({\bf r}) &=& \frac{\alpha}{4\pi}F({\bf Q}) K_0 \Big(  { r \over  \xi } \Big)\, ,
\end{eqnarray}
where  $K_0$ is the modified Bessel function, and
$ F({\bf Q}) = \cos(Q_x r_x)\cos(Q_y r_y) \ . $

Having determined the Ni induced Cu spin polarization $\langle s^z_i \rangle$,
it is straightforward to investigate the $^{17}$O NMR lineshape, determined
by the
coupling of the $I=\frac{5}{2}$ nuclear spins $^{17}{\bf I}_l$
to the Cu electron spins ${\bf s}_i$ with spatially varying mean value
$\langle s^z_i \rangle$. The hyperfine Hamiltonian is 
\begin{equation}
{\cal H}_{hf} =  \hbar^2 \gamma_n \gamma_e
 \sum_{l,i} C_{i,l} \,  {\bf s}_i\cdot ^{17}{\bf I}_l \, ,
\label{HMR}
\end{equation}
where $\gamma_n,\gamma_e$ are the gyromagnetic ratios for the $^{17}$O
nucleus and
the electron, respectively. The hyperfine coupling constants $C_{i,l}$ is
dominated
by  a  nearest  neighbor hyperfine coupling $C\approx 3.3 {\rm T}/\mu_B$
\cite{Zha96}.
However, it was recently argued that a next-nearest neighbor hyperfine
coupling
$C^\prime\approx 0.25C$ is relevant for the explanation of the spin-lattice
relaxation
rate \cite{Zha96} in La$_{ 2-x}$Sr$_x$CuO$_4$. We will therefore also
consider its
effects on the $^{17}$O NMR line. Using a mean field description of this
hyperfine
coupling  by replacing ${\bf s}_i$ by $\langle s^z_i \rangle$ of
Eq.~\ref{Sind}, we
finally obtain for the shift of the resonance at a given $^{17}$O site
${\bf r}_l$
\begin{eqnarray}
\nu_l &=& \frac{A}{T} \sum_{i,j} C_{i,l}  \chi'({\bf r}_{i}-{\bf r}_{j})\, .
\label{ox_shift}
\end{eqnarray}
Here, the sum over $i$  runs over the Cu spin sites,  coupled to the
$^{17}$O nuclear spin, whereas the $j$-summation goes over all
Ni-sites. Furthermore, the constant prefactor $A$ is given by $
\frac{5}{2}\gamma_n \gamma_e J  \hbar  C_{Curie} B_0  /( g \mu_B)^2$.
Note, $\nu_l$  as given in Eq.~\ref{ox_shift} is  the shift of the $^{17}$O
resonance  with respect to  the case  without Ni impurities.

To obtain the  $^{17}$O NMR line shape, we create a histogram
$I_o(\nu)=\sum_l\delta(\nu-\nu_l)$  counting the number of nuclei with
shift $\nu$.
Since we want to compare the resulting distribution with the experimental
data
where the line has a finite width even in the absence of impurities,
we convolute $I_o(\nu)$ with a Gaussian distribution
$\exp\left(-\nu^2/(2\sigma^2)\right)/\sqrt{2 \pi \sigma^2}$, yielding the
lineshape
$I(\nu)$. By comparison with the experiments of Ref.~\cite{BAY97} we expect
that  $\Delta \nu_0=\sqrt{2 \log 2} \sigma $ should be of the order of the
high temperature (i.e. $\xi  < 1$) Ni-impurity induced linewidth.  In the following
calculation we therefore choose $\sigma=20$kHz for both the Lorentzian and
Gaussian
$\chi'({\bf q})$. Finally, we define the resulting $\Delta \nu$ by half the
width of
the peak at half maximum.
\begin{figure} [t]
\begin{center}
\leavevmode
\epsfxsize=7.5cm
\epsffile{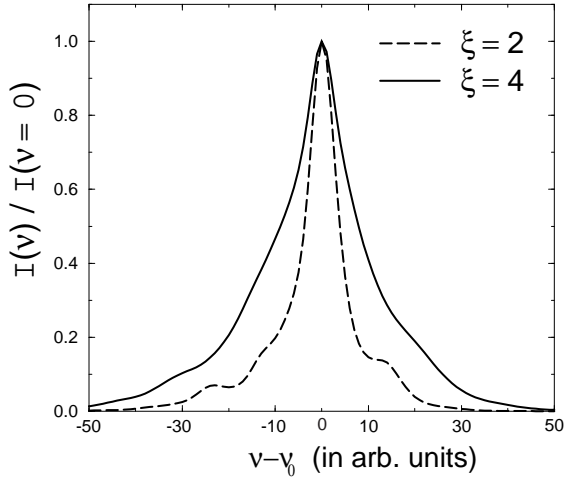}
\end{center}
\caption{The $^{17}$O lineshape $I_{conv}(\nu)/I_{conv}(\nu=0)$ for $x=2 \%$
Ni doping, $C^\prime=0$ and two different values for the correlation length
$\xi$: $\xi=2$
(dashed line), and $\xi=4$ (solid line). For both convolutions we used
$\sigma= 20 {\rm kHz}$}
\label{shape}
\end{figure}
In Fig.~\ref{shape} we present the lineshape of the $^{17}$O NMR signal,
calculated
with the Lorentzian form $\chi_{\rm L}'({\bf q})$ for two different values
of $\xi$.
We clearly observe that the line becomes broader as we increase $\xi$.  From a
comparison of Eq.(\ref{ox_shift}) with the experimentally measured
broadening we can
extract the value of the interaction $J$  in
Eq.(\ref{Hint}). 
For $C'=0$
and $\xi(200{\rm K})=4$ ($\xi(200{\rm K})=3$) we obtain $J \approx 25 \
{\rm meV}$
($43 \ {\rm meV}$). 
These values   are accompanied by some
uncertainties,  but enable us to 
 estimate  the  effects
of a Cu
spin mediated  Ni-Ni  spin (RKKY-type)  interaction.  
We find within a self consistent mean field treatment of this interaction
 that the effect of the Ni-Ni interaction
  changes $\Delta \nu$ only within a few percent, 
 consistent with the
fact that no significant deviation from a Curie law was found in
susceptibility
measurements \cite{Mah94,Men94}.

\begin{figure} [t]
\begin{center}
\leavevmode
\epsfxsize=7.5cm
\epsffile{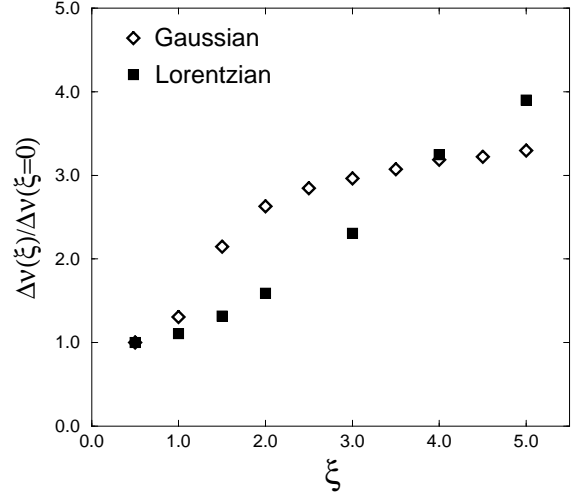}
\end{center}
\caption{$\Delta \nu(\xi)/\Delta \nu_0$ for $x=2\%$ Ni concentration
and $C^\prime=0$.
 The open diamonds present the results using $\chi_{\rm G}'({\bf q})$,
whereas the filled
 squares show the results obtained with $\chi_{\rm L}'({\bf q})$.}
\label{comp}
\end{figure}

In Fig.~\ref{comp} we present a comparison of  $\Delta \nu(\xi)$ for the
Gaussian
$\chi_{\rm G}'({\bf q})$ (open diamonds) and the Lorentzian $\chi_{\rm
L}'({\bf q})$
(filled squares).
Here we chose a Ni concentration of $x=2\%$,  $C^\prime=0$, and  ${\bf Q}=
\delta_{\rm i}
(\pm \pi, \pm \pi)$ to be incommensurate with $\delta_{\rm i}=0.94$ \cite{Mook97}. We also
compute $\Delta \nu $ for the commensurate case,  and find that in general
$\Delta \nu$ decreases.  However, since the incommensurability, $1-\delta_{\rm i}$, in
YBa$_2$Cu$_3$O$_{6.6}$, if present at all, is rather   small,
differences
are  negligible for   $\xi < 8$.  
In Fig.~\ref{comp},  we clearly see that the effect of  $\chi_{\rm G}'({\bf
q})$ and
$\chi_{\rm L}'({\bf q})$ on the behavior of the line width is {\it
qualitatively}
different. In agreement with the results by Bobroff {\it et al.} we find using
$\chi_{\rm G}'({\bf q})$ that $\Delta \nu$ is basically independent of
$\xi$ for all
physically reasonable values $2<\xi < 5$. The Lorentzian form $\chi_{\rm
L}'({\bf q})$,
however, yields a much stronger increase in $\Delta \nu$ between $\xi=2$
and $\xi=5$
than the Gaussian. This result immediately implies that a temperature
dependent $\xi$
is clearly compatible with the experimental results by Bobroff {\it et al}.
Furthermore, we find that  the function $f(\xi)$ of Eq.~\ref{dnu} behaves like
$f(\xi) \sim \xi^{3/2} $ in the Lorentzian case   and 
$f(\xi) \sim const.$ in the Gaussian case.
 This qualitatively different behavior of $ \Delta
\nu(\xi)$ for  $\chi_{\rm G}'({\bf q})$ and $\chi_{\rm L}'({\bf q})$ 
makes  this experiment extremely sensitive
to details of the momentum dependence of $\chi'({\bf q})$.

\begin{figure} [t]
\begin{center}
\leavevmode
\epsfxsize=7.5cm
\epsffile{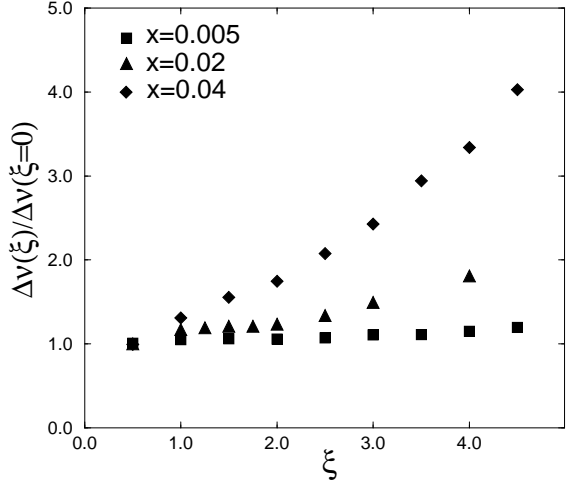}
\end{center}
\caption{$\Delta \nu(\xi)/\Delta \nu_0$ for $C^\prime=0.25C$ and three
different Ni
concentrations: $x=0.5\%$ (squares), $x=0.5\%$ (triangles) and $x=0.5\%$
(diamonds).}
\label{conc}
\end{figure}

Next we  discuss the $\xi$ dependence of $\Delta \nu$ for different values
of the Ni
concentration $x$. We present our results for a Ni concentration of $x=0.5
\%, 2 \%$
and $4 \%$ and for $C'=0.25C$ in Fig.~\ref{conc}.
In agreement with the experimental results we find that $\Delta \nu$ for a
given
$\xi$ increases with $x$. We believe that the results in Figs.~\ref{comp} and
\ref{conc} also provide an explanation for the different behavior of
$\Delta \nu$ in
the underdoped (YBa$_2$Cu$_3$O$_{6.6}$) and overdoped
(YBa$_2$Cu$_3$O$_{7}$) samples.
Bobroff {\it et al.} obtained that for the overdoped sample the variation of
$\Delta \nu$ with $T$ is much weaker than for the underdoped sample.
As far as $\chi'({\bf q})$ is concerned, the main difference between these
two regimes
consist in the value of $\xi$,  namely $\xi=1..2$ for the overdoped and
$\xi=2..4$
for the underdoped sample. We see from Figs.~\ref{comp} and \ref{conc} that
the
$\xi$ variation of $\Delta \nu$ for the overdoped sample is much weaker
than for the
underdoped one, in agreement with the experimental results.

Finally, we can use our numerical results to  investigate in more detail the
consequences of the $T$-dependence of $T \Delta \nu_{\rm imp} T_{\rm 2G}$ shown in
Fig.~\ref{prod}.
Using $g(\xi)\propto \xi$~\cite{Curro97,Pines} and the above results for $f(\xi)$,
 it follows from Eq.~\ref{product}
for the Gaussian case ($f(\xi)\sim const.$) that
 $T \Delta \nu_{\rm imp} T_{\rm 2G}  \propto \xi^{-1}$, i.e.
 $\xi$ has to increase with increasing $T$.
This result seems to be unphysical and thus strongly suggests that the
Gaussian form
$\chi_{\rm G}'({\bf q})$ is inappropriate for the description of the spin
susceptibility. On the other hand, for the Lorentzian case,
$ f(\xi) \sim \xi^{3/2}$ and  it follows 
 $T \Delta \nu_{\rm imp} T_{\rm 2G}  \propto \xi^{1/2}$, i.e. 
$\xi$ decreases as $T$ increases, as we would expect.
One can also solve Eqs.(\ref{dnu}) and (\ref{t2g}) to obtain $\alpha$ as a
function
of $T$.   However, our result  possesses error bars
which are
quite large. The conclusion that $\alpha$ is independent of $T$ is
acceptable within
those errorbars,  however, a  weak $T$  dependence cannot be excluded.

It is important to contrast  our findings with the observations of INS
experiments.
In YBa$_2$Cu$_3$O$_{6+\delta}$, INS observes a $T$-independent broad peak
around
$(\pi,\pi)$,  above   $T_c$, resembling a Gaussian
form of $\chi^{\prime \prime}({\bf q}, \omega)$~\cite{Kei,Bou}. However, strong
indications for
incommensurate peaks with Lorentzian like shape in
YBa$_2$Cu$_3$O$_{6.6}$~\cite{Mook97}
suggest that the broad structure around $(\pi,\pi)$ is only  a
superposition of
incommensurate peaks. Its  width is therefore  dominated  by the   merely
$T$-independent incommensuration instead of $\xi^{-1}$. This is consistent
with the
recent analysis by Pines~\cite{Pines} that the overall magnitude of
$\chi''({\bf q},\omega)$  in YBa$_2$Cu$_3$O$_{6+\delta}$, as obtained from NMR
experiments, necessitates a considerable improvement of the experimental
resolution
of INS experiments to resolve the incommensurate peaks in the normal state.

In conclusion we obtain from the analysis of the $^{17}$O NMR data by Bobroff
{\it et al.} and the $T_{\rm 2G}$ data by Takigawa that the correlation
length $\xi$
must have a substantial temperature dependence. A detailed analysis shows
that the
Gaussian form $\chi_{\rm G}'({\bf q})$ of the spin susceptibility can be
excluded as
an appropriate description of the spin dynamics in the doped cuprates. A
more correct
description is provided by a Lorentzian-type form $\chi_{\rm
L}'({\bf q})$,
which is fully compatible with the experimental data and a temperature
dependent $\xi$.
Though the resolution of the experiment does not allow us yet to determine the
precise $T$-dependence of $\xi$, our analysis shows that $\xi$ considerably
decreases with increasing temperature.

This work has been supported by STCS under NSF Grant No.~DMR91-20000,
the U.S. DOE Division of Materials Research under
Grant No.~DEFG02-91ER45439 (C.P.S., R.S.) and the Deutsche
Forschungsgemeinschaft (J.S.) We would like to thank H. Alloul, J. Bobroff,
A. Chubukov, D. Pines and M. Takigawa for valuable discussions.


\begin{thebibliography}{99}
\bibitem{com2} The concentration of Ni impurities in the CuO$_2$ planes
is given by $\frac{3}{2}x$, since predominantely
Cu sites in the planes are substituted by Ni.
\end{thebibliography}
\end{document}